*Article*

# Virtual Reality Training of Social Skills in Adults with Autism Spectrum Disorder: An Examination of Acceptability, Usability, User Experience, Social Skills, and Executive Functions.

**Panagiotis Kourtesis** [1,2†]*****, **Evangelia-Chrysanthi Kouklari** [3,4†], **Petros Roussos** [1], **Vasileios Mantas**[4], **Katerina Papanikolaou**[3] , **Christos Skaloumpakas**[5,6], **and Artemios Pehlivanidis**[4]

1   Department of Psychology, National and Kapodistrian University of Athens, Athens, Greece
2   Department of Psychology, University of Edinburgh, Edinburgh, United Kingdom
3   Department of Child Psychiatry, Aghia Sophia Children's Hospital, School of Medicine, National and Kapodistrian University of Athens, Athens, Greece
4   1st Department of Psychiatry, Eginition Hospital, School of Medicine, National and Kapodistrian University of Athens, Athens, Greece
5   Department of Child Psychiatry,   P. & A.Kyriakou Children's Hospital
6   Habilis, R&D Team, Athens, Greece
†   Authors share First Authorship due to equal contribution
*   Correspondence: pkourtesis@psych.uoa.gr

**Abstract:** Poor social skills in autism spectrum disorder (ASD) are associated with reduced independence in daily life. Current interventions for improving the social skills of individuals with ASD fail to represent the complexity of real-life social settings and situations. Virtual reality (VR) may facilitate social skills training in social environments and situations proximal to real life, however, more research is needed for elucidating aspects such as the acceptability, usability, and user experience of VR systems in ASD. Twenty-five participants with ASD attended a neuropsychological evaluation and three sessions of VR social skills training, incorporating 5 social scenarios with three difficulty levels for each. Participants reported high acceptability, system usability, and user experience. Significant correlations were observed between performance in social scenarios, self-reports, and executive functions. Working memory and planning ability were significant predictors of functionality level in ASD and the VR system's perceived usability respectively. Yet, performance in social scenarios was the best predictor of usability, acceptability, and functionality level. Planning ability substantially predicted performance in social scenarios, postulating an implication in social skills. Immersive VR social skills training in individuals with ASD appears an appropriate service, yet an error-less approach, which is adaptive to the individual's needs, should be preferred.

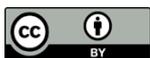

## 1. Introduction

Autism Spectrum Disorder (ASD) is a lifelong complex neurodevelopmental disorder that significantly impairs individuals' verbal and nonverbal communications, social interactions and behaviours (i.e. exhibition of restricted interests, repetitive and unusual sensory-motor behaviours) (Diagnostic and Statistical Manual of Mental Disorders [DSM]-fifth Edition; [1]). Prevalence estimates of ASD have increased over time as a recent systematic review [2] reported a global prevalence (ranging within and across regions) with a median prevalence of 100/10,000. ASD presents a striking sex difference as males are more likely to be affected relative to females (3:1 ratio) [3]. The ASD etiology is suggested to be multifactorial as both genetic and non-genetic factors (e.g., prenatal/

perinatal) may play a crucial role to the manifestation of the disorder (see [4] for a review). Since its first depiction, ASD is now regarded as a spectrum that spans from very mild to severe [5] as symptoms manifest differently in each individual based on their functionality level (level 1-requiring support, level 2-requiring substantial support; level 3-requiring very substantial support). Nevertheless, several individuals with ASD (not all) require some kind of support throughout their life [5]. Even individuals with high-functioning ASD, similar to other individuals on the mild and lower ends of the spectrum, present social skills deficits across the lifespan (up to adulthood). Adults with ASD are likely to experience problems in social and everyday life functioning due to a lack of ecological training and intervention programs during childhood and adolescence [6].

## 1.1. Social Skills and Executive Functions in ASD

Adults with ASD have been found to experience social isolation, loneliness and social anxiety (e.g., [7]) due to their deficient social skills such as atypical gaze/poor eye contact, less conversational involvement, appropriate affect, reduced verbal fluency (e.g., [8], [9]), poor understanding of social cues, and difficulties in initiating and maintaining social conversation/communication [10]. The social skills deficits in individuals with high-functioning ASD are mainly attributed to impairments in cognitive components such as Executive Functions (EF) (e.g.,[11]) or cognitive processing speed (e.g.,[12]). Indeed impaired EF is another salient characteristic of the spectrum [13] which refers to high-order, goal-directed cognitive processes that control behaviour, thought and emotions. The EF construct is seen as an umbrella term including abilities such as inhibition, working memory, and planning (not an exhaustive list; see [14] and [15] for a more detailed EF discussion). Two recent meta-analyses [16], [17] demonstrated a broad EF impairment in ASD as deficits have been consistently found in several EF aspects (e.g., inhibition, working memory, cognitive flexibility & planning) across the life span.

To implement effective interventions, research over the last decade was needed to identify which EF aspects contribute to the manifestation of social skills in ASD ([11], [18] as it is suggested that higher-order cognitive regulation is required for social interactions [19]. EF has been proposed to support the processing and manipulation of information from one's and others' perspectives, to facilitate socially interactive and communicative skills [20]. Such associations are understudied in adulthood in ASD. Limited evidence from childhood and adolescence has shown that performance-based measures of EF (e.g., auditory attention, inhibition-switching) relate to social deficits in ASD (e.g., [21], [22]) while ratings-based EF such as initiation, cognitive flexibility and working memory were found related to adaptive social skills in ASD [23], [24]. A recent study [18] also demonstrated significant associations between ratings-based EF (self-monitoring) and selective social skills (social inferencing and social knowledge) in ASD. It should be noted though that all aforementioned studies, despite their findings, did not use in vivo measures of social functioning or a naturalistic context of assessment. Social skills finally have been theoretically proposed to also depend on social cognition aspects such as mental state/emotion recognition [25] but as these aspects are not consistently associated with social impairment in ASD [26], the extent to which socio-cognitive abilities associate with the social difficulties in ASD has been debated over the years. Given these potential associations among social cognition and social skills, EF and social skills, and EF and social cognition (e.g., [27]–[29]), it has been suggested that EF may contribute to social skills both directly and indirectly [30]. Social cognition aspects are likely to partially mediate the association between EF and social skills; perhaps no single cognitive mechanism in ASD can explain the various social difficulties as argued [31], as there may be several factors potentially contributing to social skills (e.g., poor emotion regulation) that could also explain the social and behavioral problems in ASD (e.g., [32]).

## 1.2. Assessment, Training, and Intervention in ASD

Assessment of ASD impairments is critical for identifying potential difficulties and weaknesses when implementing interventions. For example, widely used measures of social functioning include the Social Responsiveness Scale (a measure of general social ability; [33], [34]), Reading the Mind in the Eyes test (a measure of mental state/emotion recognition; [35]), and the Autism Diagnostic Observation Schedule (a measure of social interaction, communication and play [36], [37]). Taking into consideration the tremendous impact of the aforementioned cognitive and social impairments on the everyday life of individuals with ASD, suitable intervention and training programs are needed [38]. Targeting cognitive deficits, cognitive training exercises in adults with ASD are usually implemented to enhance performance through repeated practice on EF tasks (e.g., [39], [40]). Cognitive training exercises encompass various intervention methods such as pen-and-paper tasks, downloadable tools, and logical games. Given the EF contribution to several aspects of social functioning, targeting specific EF aspects is thought to improve the effectiveness of training interventions in ASD [41]. However, it should be noted that cognitive training studies in ASD have been designed in recent years and thus their limited and mixed results as well as their lack of ecological validity are under ongoing discussion (e.g., [42]–[44]).

When it comes to social skills, several different strategies have been used in training and intervention programs to enhance social functioning (usually social interaction and communication) in adults with ASD. For example, strategies such as social stories and social scripts, behavioural modelling and role-playing demonstrations, video modelling and self-modelling (e.g., [45]) in the context of didactic lessons to enhance conversational skills, developing friendships, appropriate use of humour, dating, handling embarrassing feedback and peer pressure (e.g., [46]) have been used in ASD. Most psychosocial intervention and training programs in ASD however are thought to yield limited benefits [47] because of their limited ecological validity, which does not permit a generalization of the outcomes to everyday life [48], [49]. The limitations of the aforementioned methods are thought to likely arise because of the ASD literature's tendency of examining social (and/or cognitive) deficits as isolated and individual features and not by evaluating how they manifest in real-life contexts in which outcomes are influenced by relational dynamics as well [41], [50]. For that reason, computing technology with more naturalistic set-ups and role plays is a significantly effective training and intervention medium for individuals with ASD [51].

*1.3. Ecological Validity, Virtual Reality Assessments and Interventions*

Ecological validity refers to the verisimilitude (i.e., the likeness to everyday life) and veridicality (i.e., the association between the observed and real-life performance) of a neuropsychological tool, which subsequently allows the generalization to everyday life [52]. In contrast with the paper-and-pencil or computerized approaches, which incorporate static and simplistic testing and training environments and stimuli, immersive virtual reality (VR) facilitates the attainment of enhanced ecological validity and pleasantness [53]. Immersive VR neuropsychological tools may thus contribute to the understanding of everyday functionality (e.g., [54], [55]) and improve everyday physical and cognitive functioning (e.g., [56]–[58]). In the context of VR interventions in ASD, immersive VR technology facilitates the creation of simulated environments that can be used to help individuals with ASD improve social skills, communication, and behaviour [59]–[62]. These interventions aim to provide individuals with ASD with a safe and controlled environment in which to practice and develop skills, as well as to reduce anxiety and stress associated with real-world interactions [60]. VR interventions can include activities such as role-playing social scenarios, virtual social skills training, and virtual exposure therapy. However, the effective implementation of immersive VR for research and clinical purposes requires technological competence [63]. An inappropriate conceptualization of VR training may render ramifications and compromise its otherwise beneficial outcomes [62].

Nevertheless, several VR applications have efficaciously been implemented for assessment and intervention purposes. The VR Everyday Assessment Lab assesses everyday memory (prospective and episodic), attention (visuospatial and auditory), and EF (planning and multitasking), and has been found a valid and substantially more pleasant testing experience [53], which indicates everyday functionality of adults [54], [55]. The ClinicaVR: Classroom-CPT is a VR classroom that examines selective and sustained attention, and inhibition, which has been validated in children and adolescents [64]. Regarding interventions in ASD, there is preliminary evidence postulating its feasibility for being adopted in clinical and educational environments [59], [65]. Also, the use of social stories in VR has been evaluated by clinicians for implementation in clinical and educational settings for training the social skills of children with ASD [66]. Preliminary evidence suggests that VR software may improve the conversational [61], problem-solving, and communication skills of children with ASD [67]. After a VR training protocol, children with ASD showed significant improvements in emotion expression and regulation and socio-emotional reciprocity [68]. Comparably two more studies [69], [70] reported substantial enhancements in social skills in children with ASD after attending VR-based training sessions. It is important however to underline that VR interventions in ASD are still considered an emerging field and more research is needed to fully understand their efficacy, usability, the provided user experience, as well as their acceptability by individuals with ASD [59], [60], [62]. Furthermore, the relationship between performance in VR social scenarios and cognitive functioning has not yet been investigated. Finally, while there are several VR applications used in children and/or adolescents with ASD, none of the aforementioned VR applications was designed for and implemented in adults with ASD.

*1.4. VRESS*

The VR Enhancement of Social Skills (VRESS) was developed in line with the guidelines for developing VR software for research and clinical applications in the field of psychology [71], which they have been found to produce VR software that meets the criteria of the American Academy of Clinical Neuropsychology (AACN) and the National Academy of Neuropsychology (NAN) [72]. The VRESS incorporates social scenarios that are exemplary of adult activities and common in daily life, such as renewing your subscription at the gym, selecting a movie and buying a ticket at the cinema, browsing the available options and purchasing a smartphone at the phone store, attending a seminar class and interacting with the instructor and the co-students, and attending a job interview and responding to the interviewers' questions. The social scenarios were designed in line with the guidelines of Gray and Garand [73] for providing social stories that provide to individuals with ASD (i.e., the learners), a visual representation and a description of a situation or activity to prepare and instruct them on what to be expected, as well as the underlying reasons of this matter. Thus, the social scenarios of VRESS are rather descriptive than directive. The social stories were designed for individuals with ASD to comprehend and apply the intricacies of interpersonal communication to interact more appropriately and effectively. The social stories approach provides the opportunity for people with ASD to identify the context, discuss their motives, comprehend the amplifiers or the obstacles and improve their social skills [73], [74].

*1.5. Research Aims*

In sake of clarity, we provide the description of the terminology that pertains to the research aims:

- Usability: the capacity of a system to provide a condition for its users to perform the tasks safely, effectively, and efficiently while enjoying the experience.
- User Experience: how a user interacts with and experiences a product, system or service.
- Acceptability: the quality of being satisfactory and able to be agreed to or approved of being a software for a specific purpose.

This study thus aims to:

1) Evaluate the usability and user experience of an immersive VR training software of social skills (i.e., VRESS) in adults with ASD. Moreover, this study strives to

2) Examine the acceptability of the VR training software of social skills as a social service (i.e., by a service user's point of view) that may be prescribed and/or offered by clinicians, educators, and social workers to adults with ASD for training and improving their everyday social skills.

3) Investigate the relationships between cognitive functioning (i.e., aspects of social cognition and EF), independence/functionality level of individuals with ASD, performance in the VR social scenarios, as well as the acceptability, usability, and user experience ratings.

## 2. Materials and Methods

### 2.1. VRESS Scenarios and Interface

The VRESS software runs on SteamVR and is compatible with every VR headset that runs on the SteamVR platform (e.g., HTC Vive series, HTC Vive Pro series, Oculus Rift series, and Varjo VR series; see here for an exhaustive list). VRESS encompasses five social scenarios: 1) being at the gym; 2) buying a smartphone at the phone store; 3) going to the cinema to watch a movie; 4) attending a seminar class; 5) attending a job interview. Each scenario has three different difficulty levels: 1) easy; 2) moderate; and 3) difficult. Thus, the five scenarios have three diverse versions (i.e., per difficulty level), which means that there are a total of fifteen diverse scenarios in VRESS. The difficulty level is determined by the complexity of the scenario in terms of how many social tasks the users have to perform and how many 3D characters they need to interact with (e.g., just buying a ticket or discussing with friends about which film they should watch and then buying tickets for everyone). Furthermore, given that visual sensitivity (e.g., to intense light) [75] and agoraphobia and/or social phobia [76] symptoms are highly prevalent in ASD, the difficulty may further be modulated by defining the intensity of lights and the density of the population of 3D Non-Player Characters (NPCs; i.e., 3D characters that the user does not interact with them) in the virtual environment. VRESS provides a distinct User Interface (UI) to the operator's (e.g., clinician, researcher, social worker, or an educator) laptop/PC, which is not visible to the immersed user. Thus, beyond rendering the virtual environment that the user is immersed in it, VRESS provide a UI to the operator, which allows them to control the VR experience.

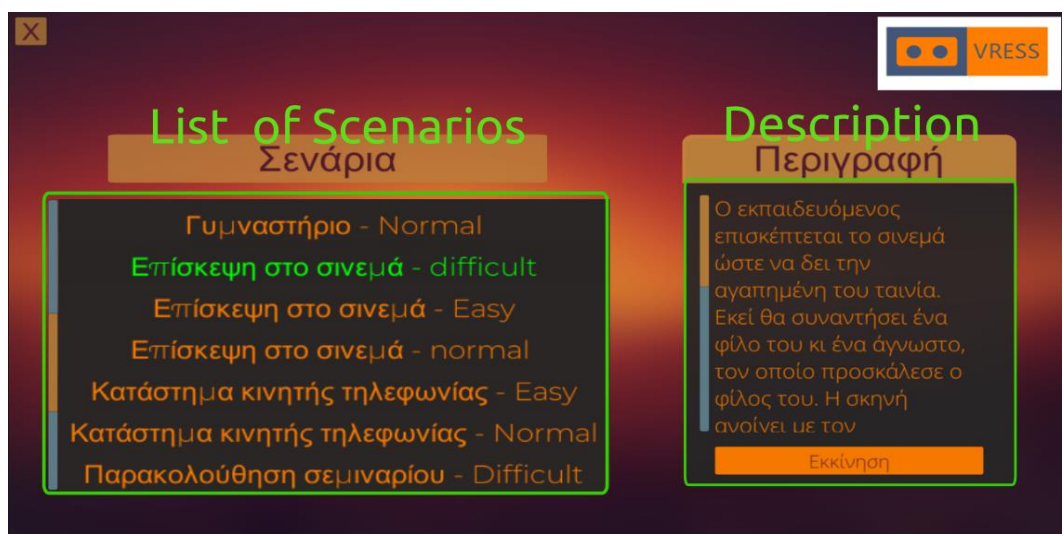

**Figure 1.** Central User Interface for Selecting the Social Scenario (Left) and Describing the Requirements to the User (Right).

There are two types of UIs. There is a central UI (see Figure 1), which appears when the VRESS application starts, that provides the operator with the available scenarios and their difficulty levels, as well as the description of each scenario level that has to be given to the users/trainees (i.e., the individuals with ASD) for understanding the social situation and the social tasks that they need to perform. This central UI also appears when the user/trainee completes a social scenario, thus, the operator requires to select and commence the next social scenario that the user/trainee has to perform. While the user/trainee is immersed in a scenario, another UI appears on the operator's screen (see Figure 2). This UI enables the operator to control which 3D character the user/trainee should interact with (note that there are 1-4 interactable 3D characters per scenario, while the other characters are just bystanders). Also, it allows the operator to manage how the interacting 3D character will respond by opting for one of the available responses. Furthermore, the operator may control the 3D character facial expressions that correspond to diverse emotional states (e.g., neutral, angry, enthusiastic, sad, happy, confused, disappointed, or surprised), as well as define the gaze direction of the 3D character (e.g., looking at the trainee, straight, or down). As mentioned above, using this UI, the operator may also control the intensity of the lighting, and the density of the NPCs' population (i.e., how many bystanders will populate the virtual environment) in the virtual environment. Finally, given that social anxiety is associated with increased heart rate [77] and the atypical eye contact [78] that are common in ASD, this UI permits the operator to monitor the user's/trainee's gaze (i.e., where the trainee is looking at, e.g., at the 3D character's torso or eyes, mouth, and nose) and heart rate.

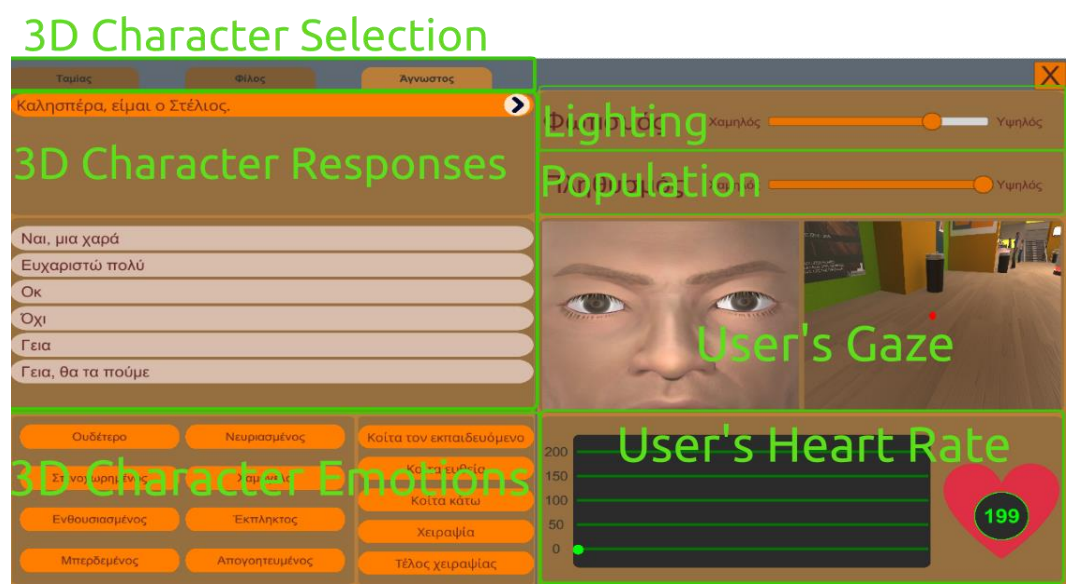

**Figure 2.** User Interface during the Scenario for Selecting the Interacting 3D Character (Top Left) and Controlling the Interacting 3D Character's Responses (Middle Left) and Emotions (Bottom Left), the Lighting's Intensity and NPC's Population Density (Top Right) in the Virtual Environment, as well as for Observing User's Gaze (Middle Right) and Heart Rate (Bottom Right).

### 2.1.1. Gym

In this scenario, the trainee is at the gym (see Figure 3). In the easy mode, the trainee has to ask the gym instructor how they may operate the running treadmill. At the medium level, the trainee has to ask another person (i.e., a co-athlete) at the gym, and then from the gym instructor how they may operate the running treadmill. Finally, at the difficult level, on top of asking around about how to operate the running treadmill, the trainee has to renew their subscription to the gym and bargain the increased fee.

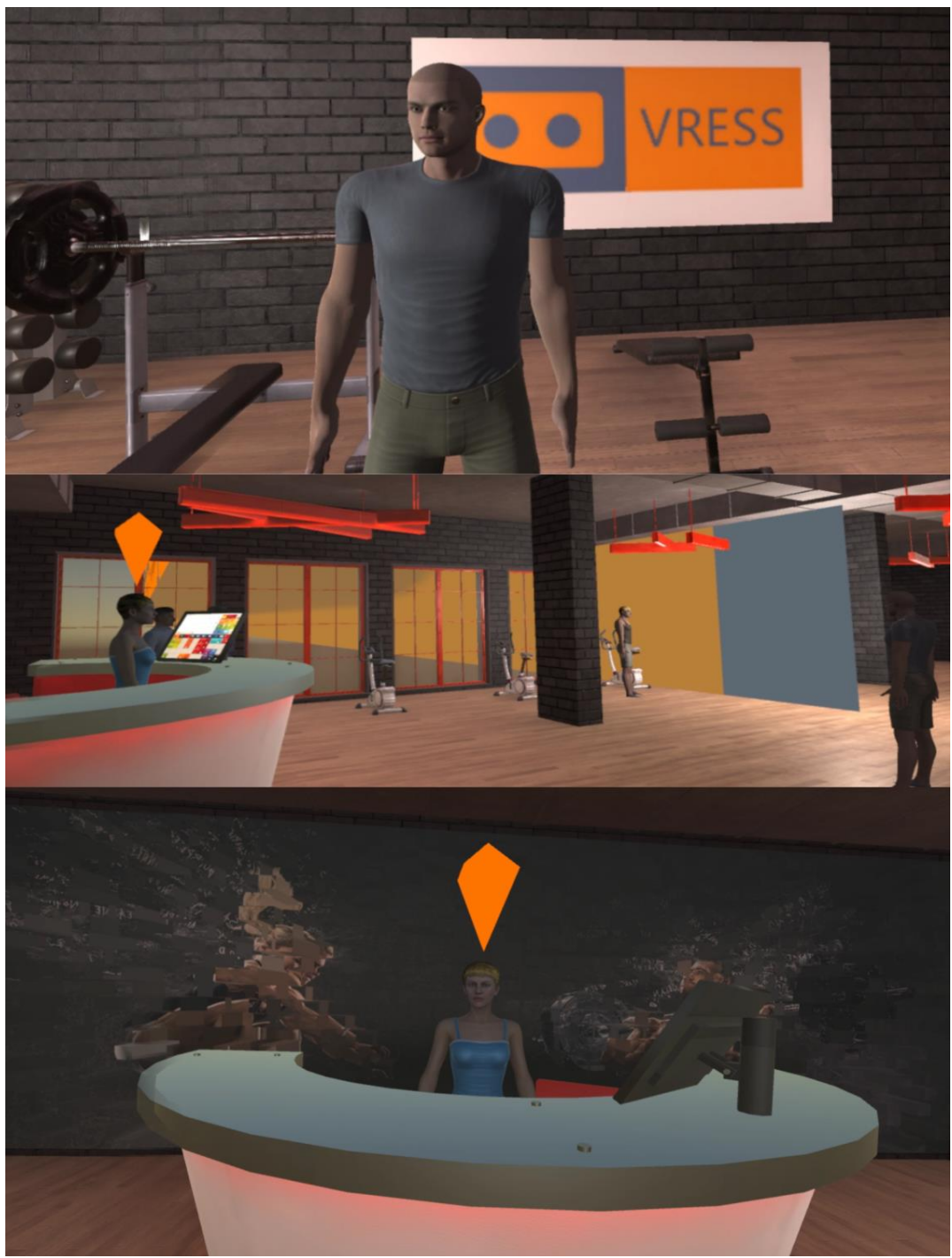

**Figure 3.** The Instructor (Top), Main Area (Middle), and Reception Desk (Bottom) of the Gym.

### 2.1.2. Phone Store

At the phone store (see Figure 4), the trainee has to buy a smartphone that costs up to €200. At the easy level, the examinee has just to browse the available options offered by the customer service person. At the moderate level, while the trainee is instructed that should buy a specific model of a brand, they have to be open to a special offer for a smartphone with better technological specifications and a lower price. At the hard level, the trainee has to browse all the available options and bargain based on a lower price that they found online.

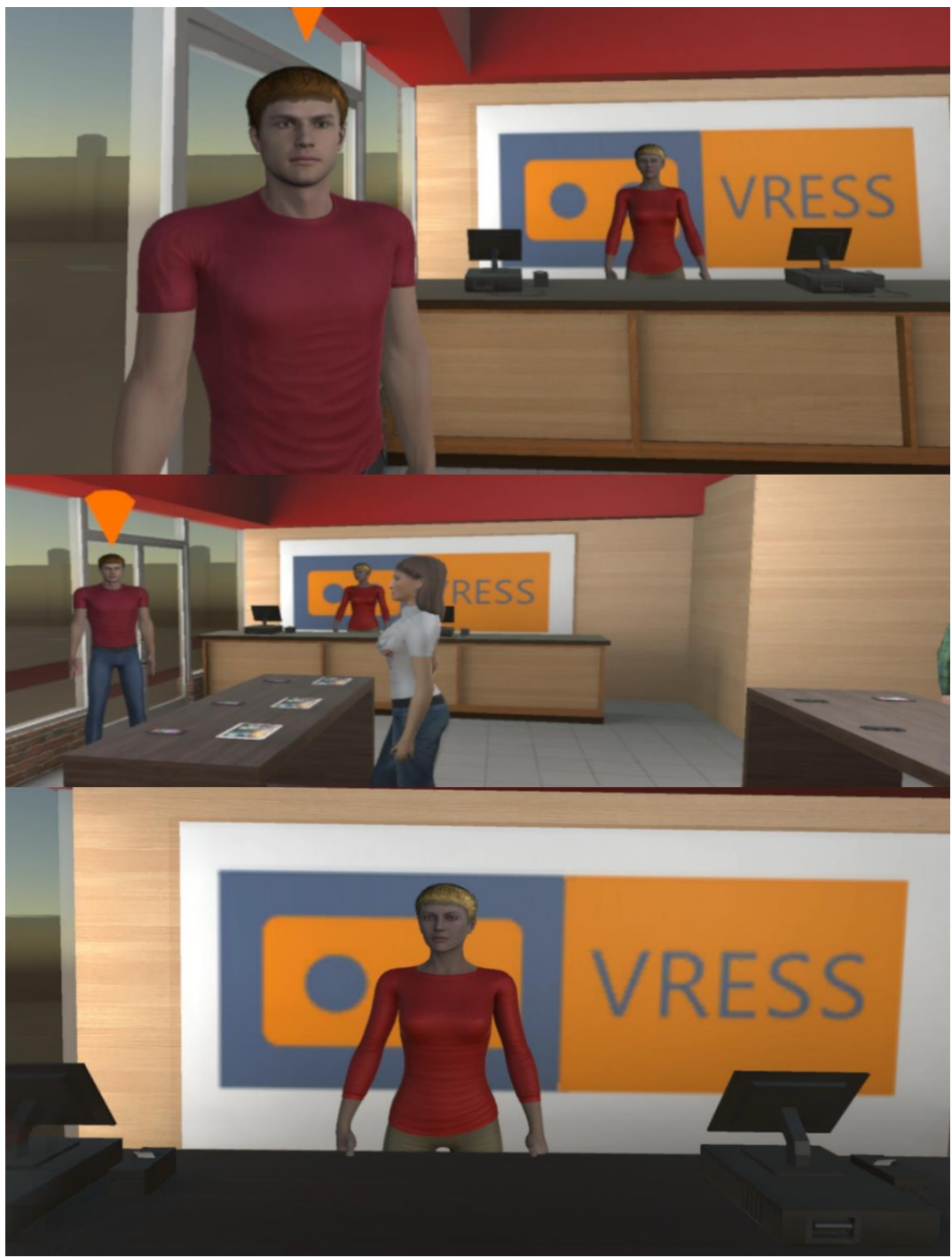

**Figure 4.** The Customer Service (Top), Main Area (Middle), and Cashier (Bottom) of the Store

### 2.1.3. Cinema

At the cinema (see Figure 5), the trainee has to select a movie and buy a ticket for this movie. At the easy level, while having a specific movie in mind, the trainee arrives late at the cinema, and they need to browse their options (e.g., next projection or another movie) and buy a ticket. At the moderate level, while having an appointment with a friend, the examinee arrives late, and they need to apologize and then buy tickets for the movie. At the difficult level, while having an appointment with a friend and another person (a friend of the friend), they need to meet them, introduce themselves, discuss finding the way to the cinema, and film genres that they like, and then choose a movie, and finally buy tickets for everybody.

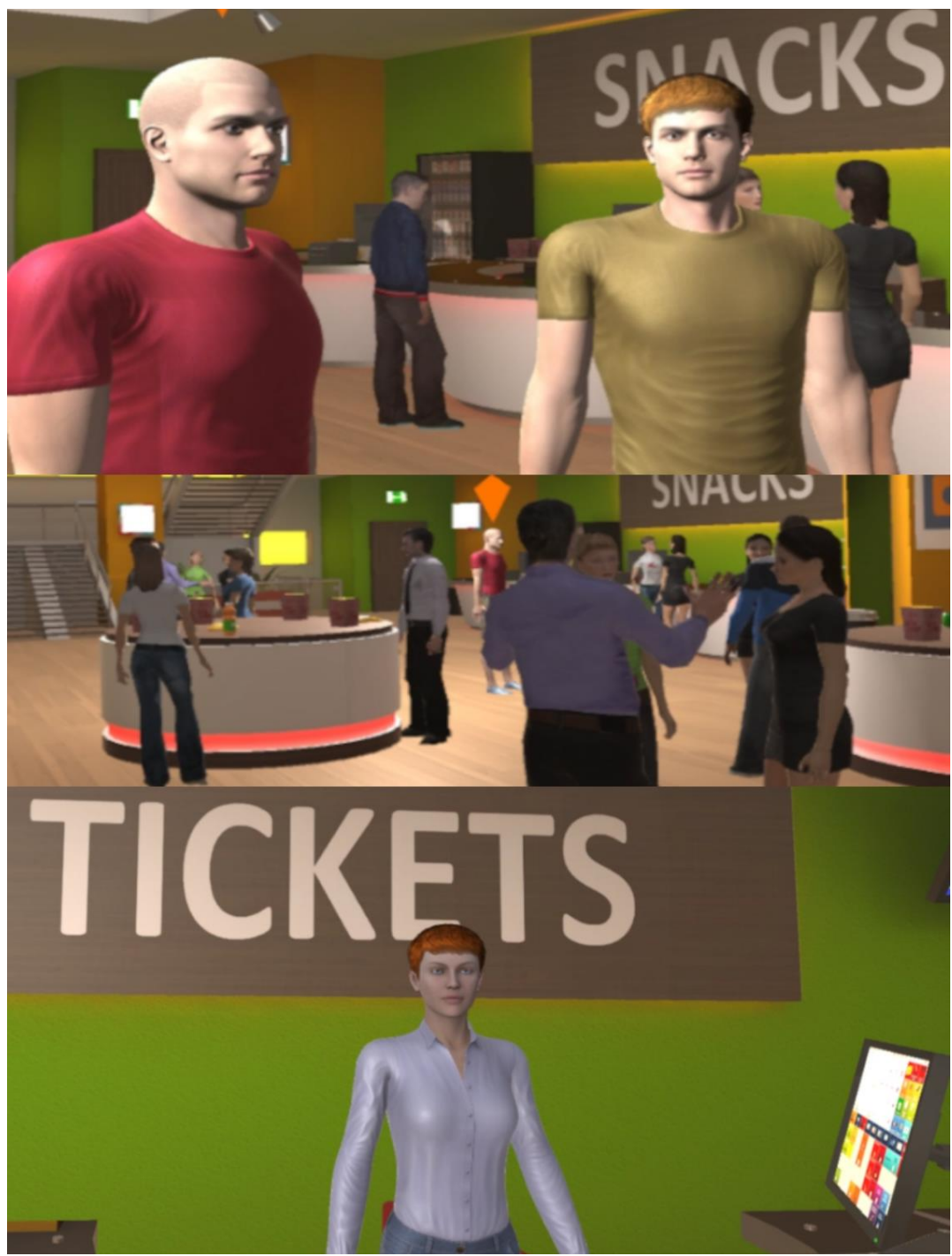

**Figure 5.** The Friends (Top), Main Area (Middle), and Tickets' Desk (Bottom) of the Cinema.

### 2.1.4. Classroom

In this scenario, the trainee has to attend a seminar class (see Figure 6). At the easy level, the trainee has to attend a 3 mins lecture by the instructor on how to find reliable information on the internet. The trainee has to respond to the instructor's question, where they have to share their opinion on Wikipedia. At the moderate level, on top of the aforementioned interaction, the trainee has to interact with their co-students during the break and ask them about their presentation. At the difficult level, the trainee has also to apologize to a co-student for making a mistake, which may undermine the reliability of their co-project and presentation.

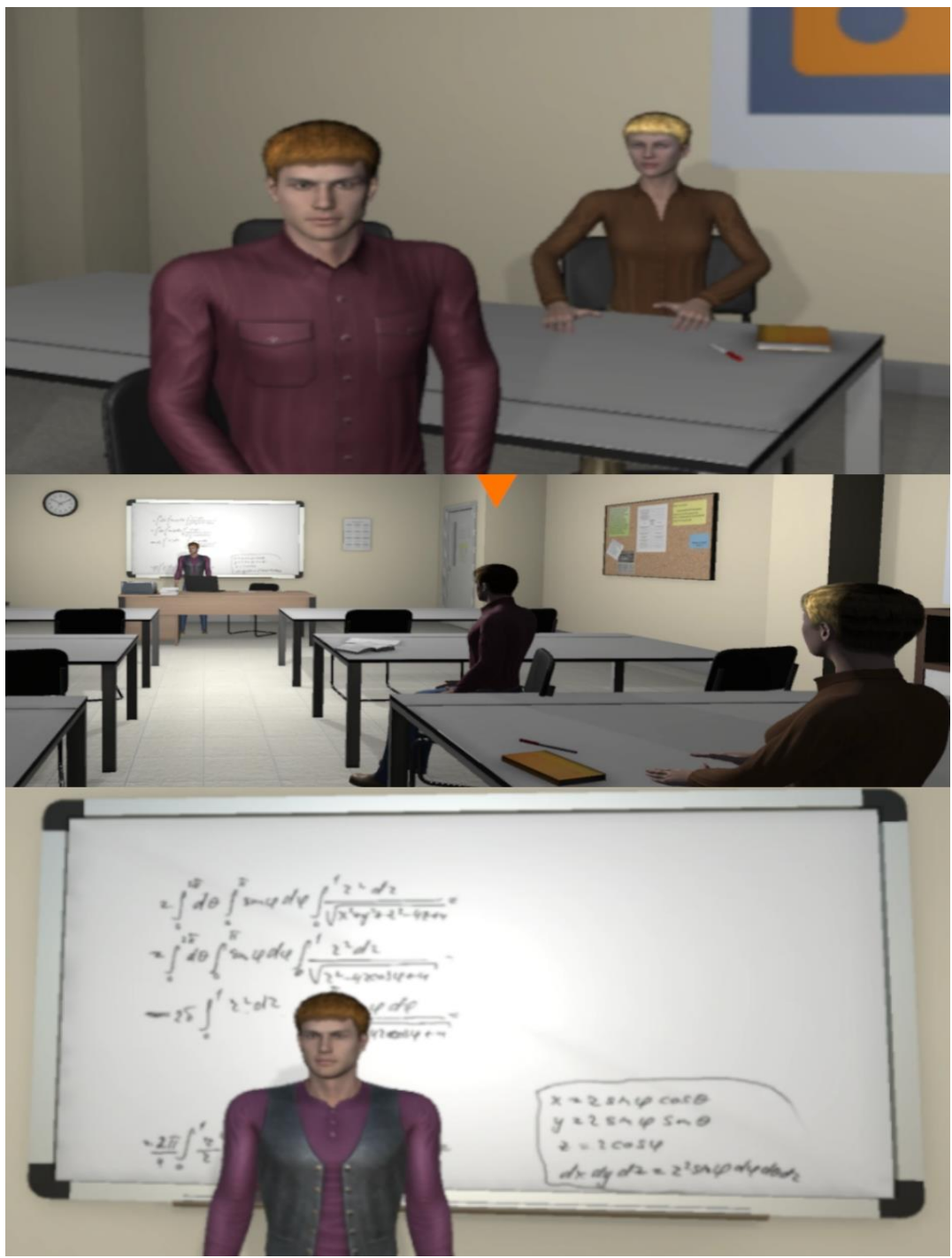

**Figure 6.** The Co-students (Top), Main Area (Middle), and Lecturer (Bottom) of the Classroom.

### 2.1.5. Interview

In this scenario, the trainee has to attend a job interview at the offices of an IT company (see Figure 7). At the easy level, the trainee is required to convince the team leader to hire them as IT assistant. At the moderate level, the trainee needs to convince both the team leader and the HR manager to hire them as IT assistant. Finally, at the difficult level, there is one more person in the waiting room, with whom, the trainee has to initiate a discussion and extract information that may assist them with getting the job. Then, the trainee has to use this information for convincing both the team leader and the HR manager that they are the best candidate for this position.

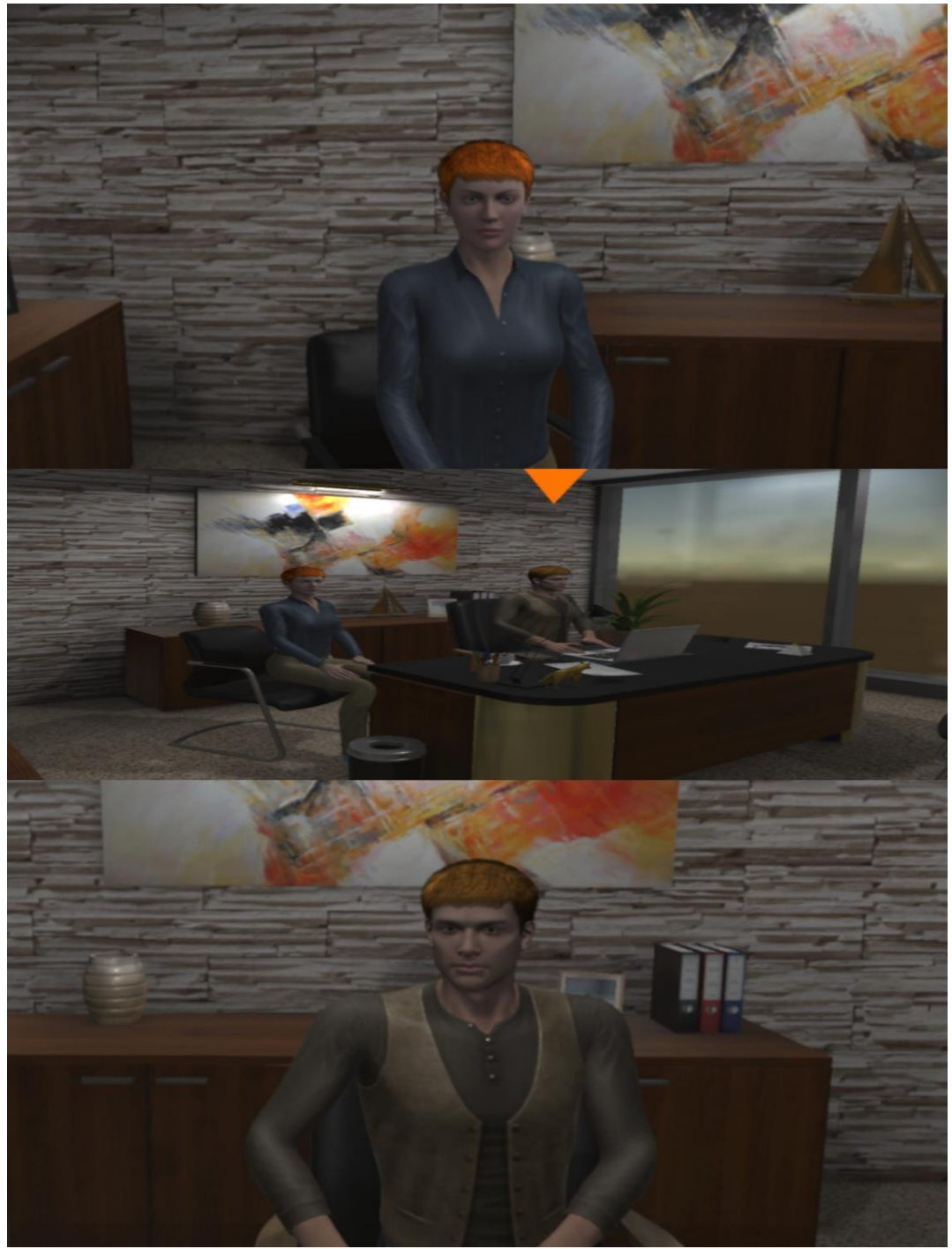

**Figure 7.** The HR Manager (Top), Main Area (Middle), and Team Leader (Bottom) of the Office.

*2.2. Neuropsychological Assessment*

2.2.1. Reading the Mind in the Eyes Test (adult version)– Mental state/emotion recognition

This test [35] measures participants' mental state/emotion recognition ability. It includes 36 pictures of the eyes (only) of different people which participants are asked to look at carefully and then choose which one of the four available options around each picture best describes what that person may be feeling/thinking. Successful performance requires participants to correctly attribute the emotional or mental state of each picture. One point was awarded for each correct answer. Scores range from 0 to 36. The Reading the Mind in the Eyes has been used in hundreds of studies to date and has been found to have good test-retest reliability [79], [80].

### 2.2.2. Tower of London - Planning

The Tower of London [81] was used to measure participants' planning skills. This test includes two identical wooden boards, one for the researcher and another for the participant. Each board has three wooden beams on which there are three wooden balls: one green, one red and one blue. Participants are asked to reproduce a series of patterns using the wooden balls only with a certain number of moves each time. Participants have to complete 12 planning problems in total: two 2-move planning problems; two 3-move planning problems; four 4-move planning problems and four 5-move planning problems. To complete all planning problems successfully, participants must follow two rules. Firstly, each planning problem must be completed in a specific number of moves and secondly, participants are allowed to remove only one ball from each beam at a time. The number of planning problems completed successfully (adhering to the rules) was recorded. One point was given for each successful completion and 0 points if participants failed. This test has been the most commonly used measure of planning across the lifespan [82], and presents good test-retest reliability [83].

### 2.2.3. Digits Recall - Verbal Working Memory

For verbal working memory, the forward and backward digit span subtests from WAIS-III [84] were administered. Participants have to recall and repeat sequences of random numbers back to the researcher in the same order (e.g., *"Please listen carefully and then repeat the following sequence of numbers back to me in the exact same order: 67893"*). Each number sequence is read at a rate of one number per second. In the backward digit span subtest, participants have to repeat the sequence of numbers in the reverse order (e.g., *"1236"* will be repeated as *"6321"*). In the case of two successfully repeated trials within each block, the examiner proceeds with the next one. Participants were awarded 1 point for each correct trial. Digit span has been extensively researched and is considered to be a highly reliable and valid measure of working memory [84], [85].

### 2.2.4. Stroop Test - Inhibition

The Stroop test [86] is a widely used measure of word-colour interference with two conditions. In the congruent condition, the colour of the ink and the printed name of the colour are similar (e.g., the colour name "yellow" is printed in yellow ink) whereas, in the incongruent condition, the colour of the ink and the printed colour word do not match. The ability to inhibit the cognitive interference occurring when the processing of a particular characteristic of a stimulus impedes the processing of a simultaneous second feature of the stimulus is known as the Stroop effect. This test assesses participants' ability to produce a contradicting response as they are asked to read the colour of the ink in which different colour words are printed, instead of reading the colour word. The response time (in seconds) was recorded. Stroop test has been found to present high test-retest reliability [87].

### 2.3. Questionnaires

### 2.3.1. Demographics and IT Skills

The participants then provided their demographic data: age in years, sex, education in years, VR experience, computing experience, and gaming experience, by responding to a custom questionnaire. VR, computing, and gaming experience were calculated by adding scores from two questions (6-item Likert Scale) for each one. The first question was regarding the participants' ability (e.g., 5 - highly skilled) to operate a VR system, a computer, or a game respectively. Comparably, the second question was pertinent to the frequency (e.g., 4 - once a week) of operating a VR system, a computer, or a game respectively. This method of providing a composite score of ability and frequency has been seen as an effective approach for evaluating the experience of an individual in using a technological medium [71], [88].

### 2.3.2. Service User Technology Acceptability Questionnaire

The Service User Technology Acceptability Questionnaire (SUTAQ) is a valid and reliable tool for evaluating the acceptability of a technological mean in a target population, which uses or will use this telehealth/telemedicine service [89]. The survey includes 22 questions, rated on a scale of 1 to 6, indicating the level of agreement with the statements provided. The survey is divided into 5 sections, each containing between 3 and 9 questions. The addition of the subscores then formulates a Total Score.

### 2.3.3. User Experience Questionnaire

The short version of the User Experience Questionnaire (UEQ) is a valid tool for evaluating the subjective opinion of users towards the user experience that a technological product facilitates [90]. The UEQ is made up of 26 items that are organized into 6 categories. Each item includes a pair of terms with opposite meanings (e.g., "efficient" and "inefficient"). Participants rate each item on a 7-point Likert scale, with responses ranging from -3 (completely agree with the negative term) to +3 (completely agree with the positive term). Half of the items begin with the positive term and the other half begin with the negative term, and they are presented in a randomized order. The addition of all the responses forms a total score, representing the overall user experience.

### 2.3.4. System Usability Scale

The System Usability Scale (SUS) is a simple and efficient tool for assessing the usability of a system [91]. It is made up of a 10-question survey that utilizes a five-point Likert scale for participant responses, ranging from "Strongly Agree" to "Strongly Disagree." The responses are combined to create a Total Score, which reflects the usability of the system [92]. The SUS can be used to evaluate a wide range of products and services, such as hardware, software, mobile devices, websites, and applications [92].

### 2.3.5. Cybersickness in Virtual Reality Questionnaire

The Cybersickness in Virtual Reality Questionnaire (CSQ-VR) is a questionnaire that evaluates the symptoms and severity of cybersickness, which has been shown a strong structural and construct validity [93], and a convergent validity against other cybersickness measurements [94]. It assesses different sub-types of cybersickness symptoms, such as nausea, disorientation, and oculomotor. It consists of 6 questions, which are presented on a 7-point Likert Scale, ranging from "1 - absent feeling" to "7 - extreme feeling". The CSQ-VR produces a Total Score, calculated by adding all the responses.

### 2.4. Participants

Twenty-five (25) adults (19 males/6 females) with an official diagnosis of ASD, aged between 19 and 52 years [*M(SD) = 29.96(9.77)*], were recruited to participate in the present study. Participants were all either high or moderate functioning (functionality levels 1 and 2 according to DSM-5; [1]), had fluent phrase speech, normal intelligence and held an official ASD diagnosis based on DSM-5 criteria [1] by psychiatrists of multidisciplinary teams with an extended clinical and research experience among adults with neurodevelopmental disorders (further see [95]–[97]). Exclusion criteria included the presence of acute psychopathology requiring urgent psychiatric treatment as well as Full Scale Intelligence Quotient (FSIQ) below 70. Ethical approval for the study was obtained by the hospital's ethics board and all participants provided the researchers with written informed consent. All participants were compensated for their participation in this project.

### 2.5. Procedures

Every participant in this study attended first a neuropsychological session, where their cognitive functioning was assessed. Also, three VR sessions were attended by the participants, where they were immersed and performed the VR social scenarios per difficulty level. At the end of the last VR session, they responded to the questionnaires.

### 2.5.1. Neuropsychological Session

Participants were assessed in the mental state/emotion recognition test and EF measures across one appointment (60 mins) by a researcher. During the neuropsychological assessment, the mental state/emotion recognition test was addressed first whereas the order of the EF tasks was randomised across participants. Participants' responses were scored at the end of each session. Breaks were included when necessary.

2.5.2. VR Sessions

Participants were immersed in the VRESS by using an HTC Vive Pro Eye headset, which substantially exceeds the recommended hardware criteria for avoiding or alleviating any cybersickness symptomatology [63]. HTC Vive Pro Eye integrates an eye-tracker with a 120Hz refresh rate and a tracking accuracy of 0.5°-1.1°. There were three VR sessions per participant, corresponding to the three difficulty levels: easy, moderate, and difficult. Five scenarios were performed in each session. The order of the five scenarios was counterbalanced between the participants (i.e., a complete counterbalance was achieved for every five participants). The order of the scenarios was then the same for the participant in each difficulty-level session. The three different sessions had a week gap between them. At the beginning of every VR session, a demonstration of how to properly use and handle the headset and controllers was provided to all participants. The participants performed the social scenarios in a standing or a sitting position (see Figure 8), respectively to the scenario's requirement (e.g., classroom and interview required a sitting position). At the end of the third VR session, the participants responded to the questionnaires (see subsection 2.3 above), followed by a debriefing session, where the research aims were explained to them.

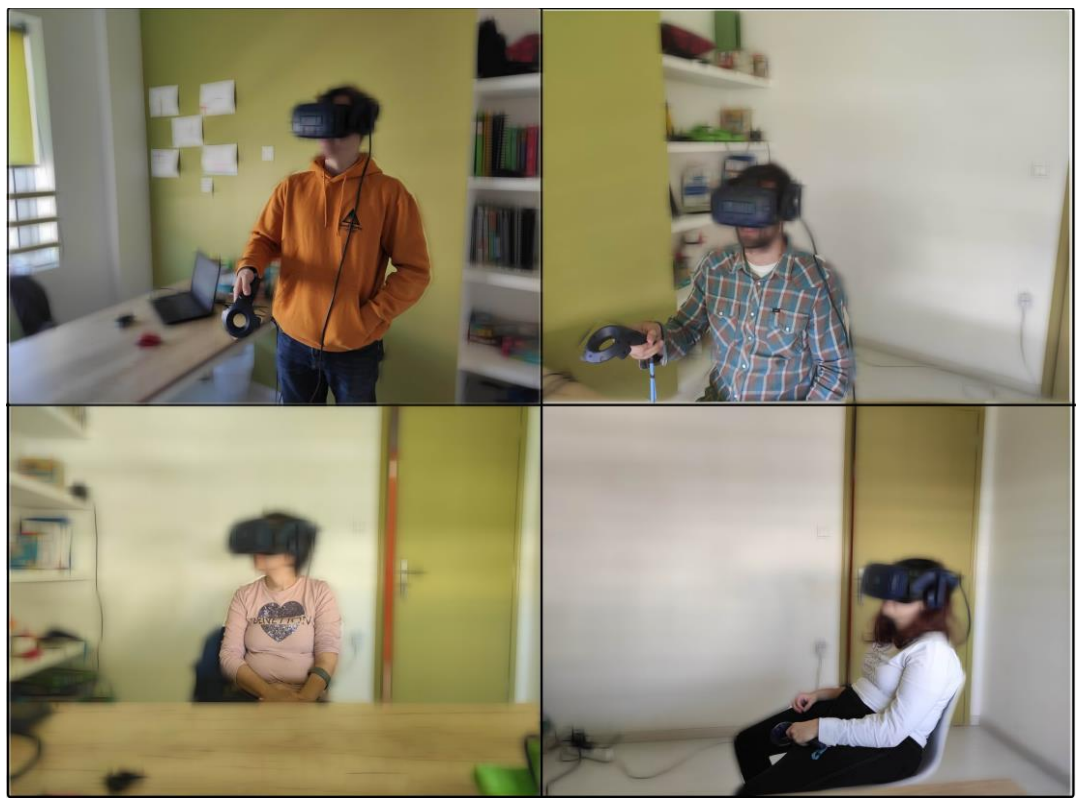

**Figure 8.** Participants Performing the VR Social Scenarios in a Standing or Sitting Position. Images are blurred to prevent the identification of participants.

2.5.3 Performance Evaluation in the VR Social Scenarios

The researcher who conducted the VR sessions was also scoring the performance of the participants. As mentioned above (see subsection 2.1), the researcher, who was the operator of the VRESS, controlled the VR experience. During each scenario, the participants had to interact with the 3D characters in the virtual environment by simply talking to them. The operator of VRESS (i.e., the researcher), then was choosing the response of the 3D character. The response was the most appropriate to what the participant said to the 3D character. In case of an inappropriate social interaction by the participant (e.g., saying something irrelevant, being silent, repeating the same thing, or making a faux pas), the operator was providing a prompt to the participant to assist them with reacting appropriately. If the participant was again not behaving in consistency with the social situation's demands, then the operator was opting for a response for the 3D character that will continue the social scenario's storyline. The performance of the participants was evaluated by two overall scores, the *Task Completion Score* and the *Prompts Score*. The task completion was calculated by the number of social tasks/interactions that were correctly performed in each social scenario. The participants received 2 points when they performed all the social tasks/interactions efficiently, 1 point when they performed half or more than the half social interactions correctly, and 0 points when they performed appropriately less than half social interactions.   A total score for task completion was calculated per difficulty level (i.e., the addition of all the points gathered in the five social scenarios of this difficulty level). An overall task completion score was formed by adding the total scores per difficulty level.   Equally, a prompts' score was calculated. The number of prompts, which were given to the participants in each social scenario, was noted. The addition of all prompts per difficulty level formulated a corresponding total score for each difficulty level. The overall prompts' score was formulated by the sum of all total scores of each difficulty level.

*2.6. Statistical Analyses*

Descriptive statistical analysis was performed to provide an overview of the sample. Pearson's correlational analyses were performed to investigate the relationships between cognitive functions, performance in VR social scenarios, and acceptability, usability, and user experience ratings. Kendal's Tau correlational analyses were conducted to inspect the associations with the functionality level of individuals with ASD (i.e., 1- high-functioning and 2- moderate functioning; dichotomous variable). Generalized regression analyses were performed to inspect the ability of the performance variables to predict the functionality level of individuals with ASD. Linear regression analyses were used to examine the predictors of acceptability, usability, user experience, and the number of prompts.   The *R*   language [98] on R Studio [99] was used for performing the analyses. The *best-Normalize* R package [100] was used to transform and centralize the data since the continuous variables violated the normality assumption. The distribution of the continuous data was then normal. For performing the respective analyses, the *psych* (correlational analyses) [101], the *ggplot2* (plots) [102], and the *stats* (regression analyses) [98] R packages were used.

## 3. Results

*3.1. Descriptive Statistics*

3.1.1. Demographic Information

The descriptive statistics of the population are displayed in Table 1. The age of participants seems to extend to the whole spectrum of early adulthood (i.e., 20-39 years) and the early half of middle adulthood (i.e., 40-59 years) albeit the population is predominantly representative of the former. The education level of the participants indicates that the majority had a university (undergraduate), college, or professional post-high school education. Furthermore, participants experienced none to very mild cybersickness symptoms, postulating that the cybersickness did not interfere with performance or user experience metrics. The VR experience of the population was relatively low. However, the computing experience appears to be on the upper tier of the possible scores, indicating that the participants were experienced in using computers in their daily life. Yet, the gaming experience was balanced, suggesting that the sample consisted equally of both gamers and non-gamers.

3.1.2. Performance on Neuropsychological Tests and Social Scenarios

Regarding the performance on the social scenarios of VRESS, the descriptive statistics for the task completion score indicate a ceiling effect (i.e., the vast majority of participants received a high score, which is close to or exactly at the maximum possible score) and a limited variance. On the other hand, the number of prompts required to efficaciously perform the social interactions in every scenario appears to have a greater range and variance, postulating that it can be a better discriminator of the performance differences among participants. Finally, regarding the performance on neuropsychological tests, the descriptive statistics indicate an intermediate (e.g., Digit Span Backward) or upper intermediate performance on emotional recognition and EF tests. However, the correct responses on the Stroop test reveal a ceiling effect, while the participants' response times show a greater variance and range on this test.

**Table 1.** Descriptive Statistics of the Sample

| Variables | Mean (SD) | Range | Maximum Score |
|---|---|---|---|
| Sex (Female/Male) | 6/19 | - | - |
| ASD Functionality Level (1/2) | 14/11 | - | - |
| Age | 29.96 (9.76) | 19 – 52 | - |
| Education | 15.88 (2.26) | 12 – 20 | - |
| Cybersickness | 7.52 (2.04) | 6 – 14 | 42 |
| VR Experience | 3.48 (1.38) | 2 – 6 | 12 |
| Computing Experience | 8.96 (2.38) | 3 – 12 | 12 |
| Gaming Experience | 6.68 (3.13) | 2 – 12 | 12 |
| Acceptability | 104.28 (21.07) | 49 – 127 | 132 |
| User Experience | 126.00 (26.33) | 78 – 180 | 180 |
| Usability | 77.12 (12.08) | 54 – 98 | 100 |
| Task Completion Score | 27.56 (1.82) | 24 – 30 | 30 |
| Prompts' Score | 11.68 (4.38) | 6 – 20 | - |
| RTMIE | 25.72 (5.21) | 8 – 33 | 36 |
| Digit Span Forward | 9.88 (2.33) | 4 – 14 | 16 |
| Digit Span Backward | 7.52 (2.88) | 2 – 13 | 14 |
| Tower of London | 7.88 (2.06) | 3 – 11 | 12 |
| Stroop – Correct Responses | 48.24 (4.01) | 30 – 50 | 50 |
| Strop – Response Time* | 65.44 (24.99) | 36 – 159 | - |

RTMIE = Reading The Mind in the Eyes; *measured in seconds

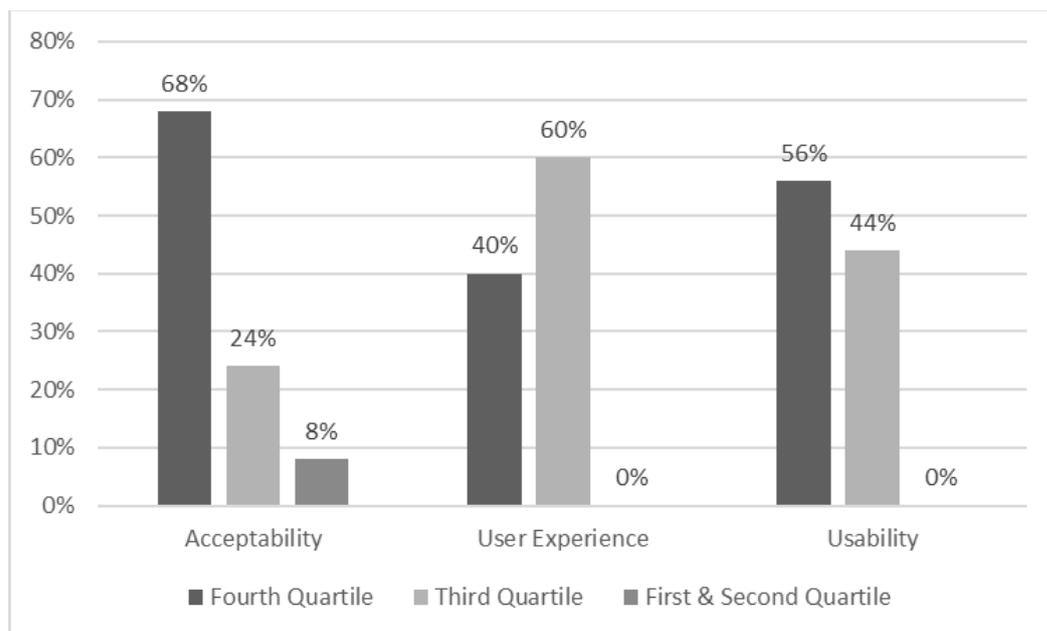

**Figure 9.** Percentages of the Responses per Scores' Quartile; Quartiles were defined by the maximum possible score of each questionnaire divided by four. Fourth quartile = highest possible scores, first quartile = lowest possible scores.

### 3.1.3. Acceptability, User Experience, and Usability Ratings

As Table 1 and Figure 9 illustrate, the vast majority of participants reported very high acceptability of using immersive VR training as a social/health service. 68% of the responses were at the highest quartile postulating substantially high acceptability. Notably, 92% of the participants' responses had an overall score above the medium scores of SUTAQ (i.e., 66), which indicates a very high rate of acceptability [103]. For user experience, the majority of responses were in the third quartile (see Figure 9), while all the responses were above the medium score. These scores postulate a high to very high user experience [90], [104]. Comparably, 100% of the respondents gave scores in the third and fourth quartile of the possible scores (see Figure 9), which indicates good to excellent usability [92]. Likewise, considering both the mean and standard deviation of the usability scores (see Table 1), the usability score of VRESS postulates a very good to excellent usability rating [92].

### 3.2. *Pearson's and Kendall's Tau Correlations*

#### 3.2.1. Demographics Correlations with Self-Reports and Performance

Demographic information of participants showed no significant associations with the acceptability, usability, and user experience ratings (see Table 2); however, significant correlations were observed with the performance on social scenarios and neuropsychological tests (see Table 3). Specifically, the participant's age was positively correlated with the correct responses on the Stroop test, yet no other correlations were detected. Similarly, the educational level of the participants revealed positive associations only with the Digit Span scores, Forward and Backward recall, respectively. Participants' experience in using VR systems showed no significant correlations with any of the performance metrics. However, both computing and gaming experience were substantially correlated with the performance on RTMIE, postulating that individuals with higher experience in using computers and/or playing video games are better at recognizing the emotional/mental states of others. In line with this finding, computing experience was also associated with the overall task completion score in VR social scenarios. Finally, the experience of playing video games was negatively associated with response time in the Stroop task.

**Table 2.** Pearson's Correlations between Demographics and Self-Reports

|  |  | Age | Education | VR XP | Computing XP | Gaming XP |
|---|---|---|---|---|---|---|
| **Acceptability** | Pearson's r | 0.345 | -0.044 | 0.071 | 0.213 | -0.141 |
|  | p-value | 0.091 | 0.834 | 0.736 | 0.306 | 0.503 |
| **User Experience** | Pearson's r | 0.351 | -0.340 | -0.061 | 0.096 | -0.183 |
|  | p-value | 0.085 | 0.096 | 0.771 | 0.647 | 0.382 |
| **Usability** | Pearson's r | 0.119 | 0.031 | 0.269 | 0.310 | 0.169 |
|  | p-value | 0.572 | 0.884 | 0.193 | 0.131 | 0.420 |

XP = Experience; *p <.05, ** p <.01, ***p <.001.

**Table 3.** Pearson's Correlations between Demographics and Performance Metrics

|  |  | Age | Education | VR XP | Computing XP | Gaming XP |
|---|---|---|---|---|---|---|
| **RTMIE** | Pearson's r | 0.059 | 0.372 | 0.276 | *0.427** | *0.503** |
|  | p-value | 0.780 | 0.067 | 0.181 | *0.033* | *0.010* |
| **DS Forward** | Pearson's r | -0.064 | *0.412** | 0.281 | 0.331 | 0.348 |
|  | p-value | 0.760 | *0.040* | 0.173 | 0.106 | 0.088 |
| **DS Backward** | Pearson's r | 0.152 | *0.413** | 0.108 | 0.195 | 0.237 |
|  | p-value | 0.469 | *0.040* | 0.607 | 0.349 | 0.255 |
| **ToL** | Pearson's r | 0.206 | 0.349 | 0.356 | 0.393 | 0.349 |
|  | p-value | 0.323 | 0.088 | 0.081 | 0.052 | 0.087 |
| **Stroop CR** | Pearson's r | *0.411** | -0.049 | 0.100 | 0.267 | 0.193 |
|  | p-value | *0.041* | 0.815 | 0.635 | 0.197 | 0.354 |
| **Stroop RT** | Pearson's r | 0.037 | -0.227 | -0.340 | -0.380 | *-0.483** |
|  | p-value | 0.860 | 0.276 | 0.097 | 0.061 | *0.015* |
| **Prompts** | Pearson's r | -0.064 | -0.096 | 0.347 | -0.169 | -0.115 |
|  | p-value | 0.760 | 0.647 | 0.059 | 0.419 | 0.585 |
| **Task Completion** | Pearson's r | 0.177 | 0.206 | -0.392 | *0.468** | 0.196 |
|  | p-value | 0.396 | 0.324 | 0.053 | *0.018* | 0.349 |

XP = Experience; *p <.05, ** p <.01, ***p <.001.

### 3.2.2. Self-Reports, Performance Metrics, and ASD Functionality Level

The functionality level of individuals with ASD revealed significant correlations only with the usability rating, the number of prompts required to perform the social scenarios, the performance on Digit Span Forward recall, and the response time on the Stroop test (see Table 4). Specifically, a higher functionality level is associated with higher ratings in the system's perceived usability, requiring fewer prompts for performing social tasks, having a greater verbal working memory span and faster inhibition. Moreover, substantial positive associations were detected between acceptability, usability, and user experience (see Table 5), postulating that higher usability of a VR system facilitates a better user experience and increased acceptability as a digital social/health service.

**Table 4.** Kendall's Tau Significant Correlations with ASD Functionality Level

|  |  | Usability | Prompts | DS Forward | Stroop RT |
|---|---|---|---|---|---|
| **ASD Functionality Level** | Kendall's Tau B | *0.488*** | *-0.406** | *0.416** | *-0.365** |
|  | p-value | *0.005* | *0.021* | *0.021* | *0.033* |

DS = Digit Span; RT = Response time; *p <.05, ** p <.01, ***p <.001.

Furthermore, performance in social scenarios and performance on neuropsychological tests were significantly correlated with self-reports on acceptability, usability, and user experience. Requiring more prompts to perform the social interactions in the scenarios was associated with lower acceptability and the system's perceived usability, as well as with less social tasks completion. Equally, a larger task completion score was correlated to a higher system's perceived usability. Usability also revealed positive correlations with both Digit Span scores (i.e., Forward and Backward recall; greater working memory span) and Tower of London (i.e., better planning ability), and a negative correlation with the response time in the Stroop test (i.e., faster inhibition). Finally, the number of prompts required for performing the social tasks showed substantial negative associations with the Digit Span Forward recall and the Tower of London, postulating that greater working memory and planning ability respectively assist with performing social interactions without requiring support and/or reminders.

**Table 5.** Pearson's Correlations between Self-Reports and Performance Metrics

| | | Acceptability | User Experience | Usability | Prompts | Task Completion |
|---|---|---|---|---|---|---|
| **Acceptability** | Pearson's r | - | - | - | - | - |
| | p-value | - | - | - | - | - |
| **User Experience** | Pearson's r | *0.534\*\** | - | - | - | - |
| | p-value | *0.006* | - | - | - | - |
| **Usability** | Pearson's r | *0.693\*\*\** | *0.486\** | - | - | - |
| | p-value | *< .001* | *0.014* | - | - | - |
| **Prompts** | Pearson's r | *-0.451\** | -0.200 | *-0.757\*\*\** | - | - |
| | p-value | *0.024* | 0.339 | *< .001* | - | - |
| **Task Completion** | Pearson's r | 0.366 | 0.272 | *0.523\*\** | *-0.635\*\*\** | - |
| | p-value | 0.072 | 0.189 | *0.007* | *< .001* | - |
| **RTMIE** | Pearson's r | -0.076 | -0.158 | 0.004 | -0.014 | 0.107 |
| | p-value | 0.716 | 0.452 | 0.987 | 0.947 | 0.611 |
| **DS Forward** | Pearson's r | 0.387 | 0.004 | *0.628\*\*\** | *-0.452\** | 0.285 |
| | p-value | 0.056 | 0.986 | *< .001* | *0.023* | 0.167 |
| **DS Backward** | Pearson's r | 0.228 | 0.072 | *0.477\*\** | -0.299 | 0.207 |
| | p-value | 0.273 | 0.733 | *0.016* | 0.146 | 0.321 |
| **ToL** | Pearson's r | 0.354 | 0.001 | *0.685\*\*\** | *-0.499\** | 0.262 |
| | p-value | 0.083 | 0.995 | *< .001* | *0.011* | 0.206 |
| **Stroop CR** | Pearson's r | 0.039 | 0.145 | 0.182 | -0.187 | 0.370 |
| | p-value | 0.852 | 0.490 | 0.383 | 0.370 | 0.069 |
| **Stroop RT** | Pearson's r | -0.203 | 0.032 | *-0.569\*\** | 0.313 | 0.118 |
| | p-value | 0.330 | 0.879 | *0.003* | 0.128 | 0.576 |

RTMIE = Reading the mind in the eyes test.; DS = Digit Span; ToL = Tower of London; CR = Correct responses; RT = Response time; *p <.05, ** p <.01, ***p <.001.

*3.3. Linear Regression and Generalized Linear Models*

3.3.1. ASD Functionality Level

Three models were found for predicting the functionality level of individuals with ASD (see Table 6). All models were significantly better than the null model. The models showed high $R^2$, indicating that they explain 26% - 30% of the variance of functionality level. While all predictors showed a large β coefficient, the number of prompts had the highest one, suggesting that requiring more prompts substantially predicts a lower functionality level (see Figure 10). Similarly, reduced working memory capacity and slower inhibition respectively predict a lower functionality level in ASD.

**Table 6.** Best Generalized Linear Models For Predicting ASD Functionality Level.

| Predictor | $\chi^2$ | p-value ($\chi^2$) | β coefficient | p-value (β) | R² |
|---|---|---|---|---|---|
| Prompts | 6.22 | 0.01* | -1.25 | 0.03* | 0.30 |
| DS Forward | 5.83 | 0.02* | 1.22 | 0.04* | 0.28 |
| Stroop RT | 5.30 | 0.02* | -1.09 | 0.04* | 0.26 |

DS = Digit Span; RT = Response time; *p <.05, ** p <.01, ***p <.001.

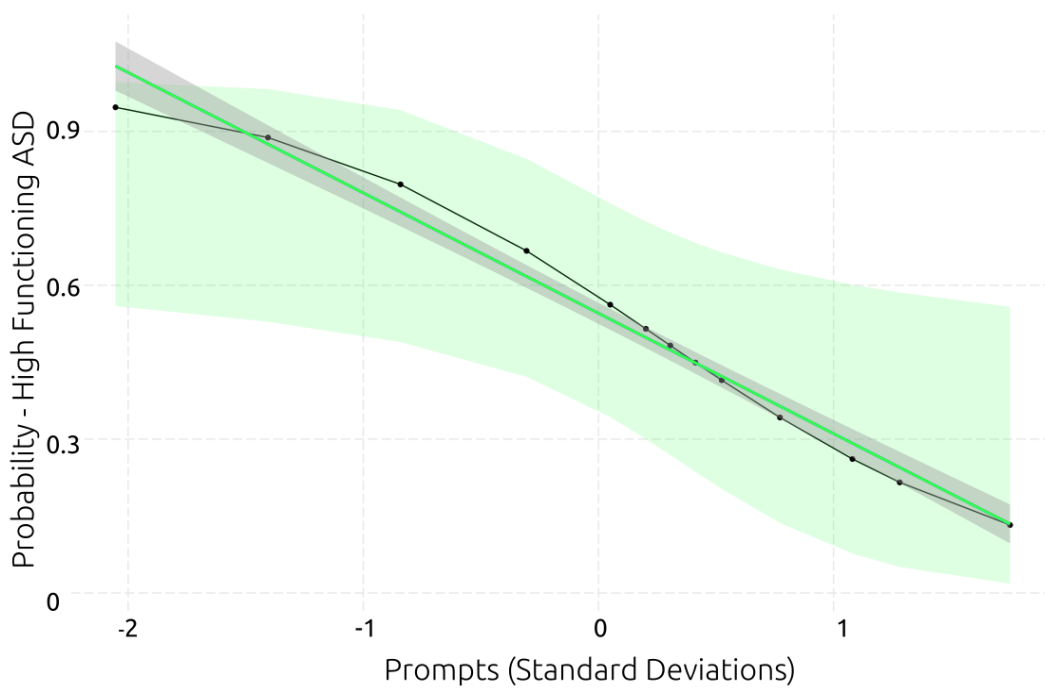

**Figure 10.** Best Generalized Linear Model For Predicting ASD Functionality Level.

### 3.3.2. Performance in VR Social Scenarios

Considering that the task completion showed a ceiling effect and a reduced range of scores and variance (see subsubsection 3.1.2), while the number of prompts did not suffer from a ceiling effect and had a long-range and rich variance of scores, the number of prompts was preferred as an indicator of performance on VR social scenarios. Only the Digit Span Forward and the Tower of London were significant predictors of the number of prompts. The model with the Digit Span Forward as a predictor was significantly better than the null model, explained the 20% of the variance of the number of prompts, and had a relatively large β coefficient [$F(1,23) = 5.91$, $p = 0.02$, $R^2 = 0.20$; $β = -0.46$, $p = 0.02$], suggesting that the lower verbal working memory span predicts that a greater number of prompts is required for performing efficiently the social tasks in the VR scenarios. However, the score on the Tower of London was the best predictor of the number of prompts (see Figure 11). The model showed that the planning ability explains 25% of the variance of the number of prompts, and had a slightly larger β coefficient, which indicates that higher planning ability predicts that is required a smaller number of prompts for efficiently interacting and completing the VR social scenarios.

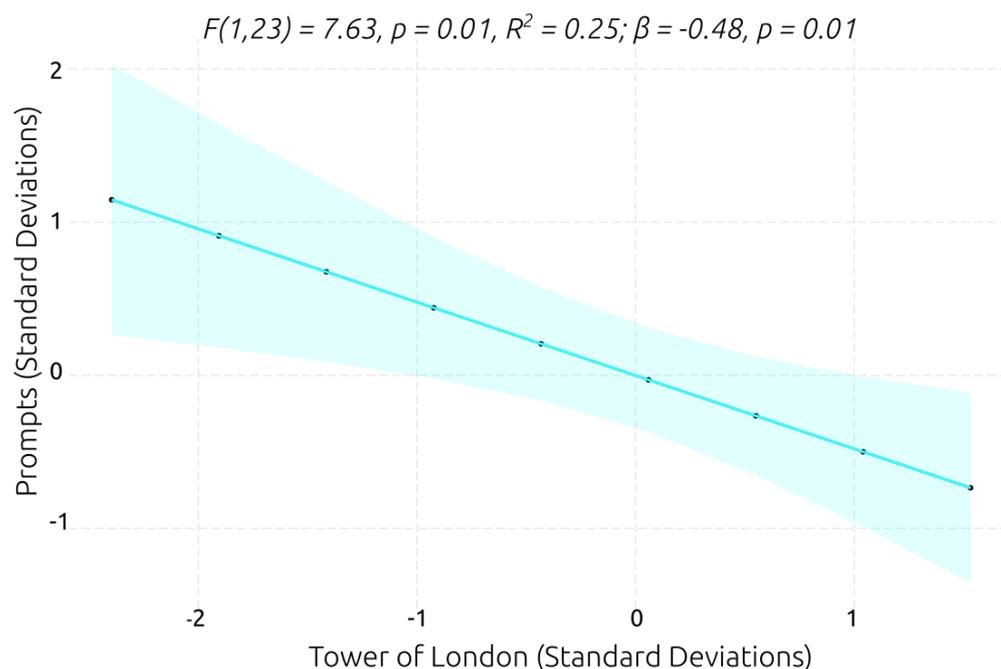

**Figure 11.** Best Linear Regression Model For Predicting Prompts' Number.

### 3.3.3. Service User's Acceptability and User Experience

The model with the number of prompts as a predictor of service users' acceptability was the only one that was significantly better than the null model (see Figure 12). The model showed that the number of prompts explained 22% of the variance of acceptability ratings, and had a relatively large β coefficient, which indicates that individuals with ASD, who required more prompts for performing the social scenarios, provided lower acceptability ratings. Comparably, the only model for predicting user experience ratings, that was substantially better than the null model, was the one with the system's perceived usability rating as a predictor (see Figure 13). The model explained the 25% variance of user's experience rating, and had a large β coefficient, postulating that the individuals who perceived that the VR system has higher usability, reported a higher user experience.

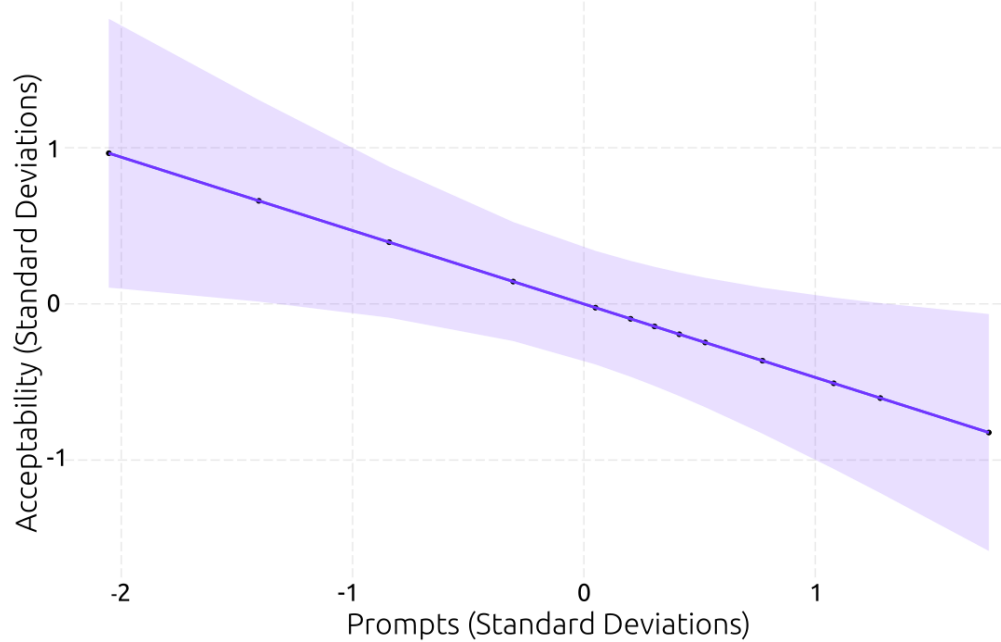

**Figure 12.** Best Linear Regression Model For Predicting Service User's Acceptability.

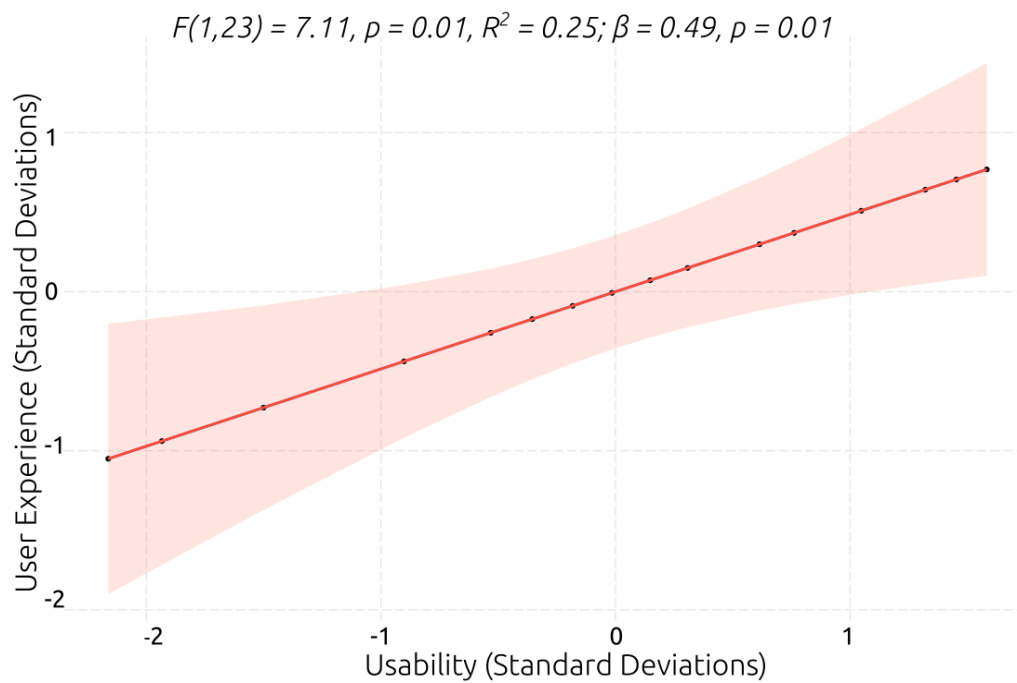

$F(1,23) = 7.11, p = 0.01, R^2 = 0.25; \beta = 0.49, p = 0.01$

**Figure 13.** Best Linear Regression Model For Predicting User's Experience.

### 3.3.4. System's Perceived Usability

For predicting the system's perceived usability, four models with a single predictor were detected, which were significantly better than the null model (see Table 7). All predictors had a large (e.g., task completion) to a very large (e.g., number of prompts) β coefficient, postulating that individuals with a better working memory capacity, planning ability, and/or performance on VR social scenarios, perceived a higher system's usability. The overall task completion in the VR social scenarios, the Digit Span Forward recall (i.e., verbal working memory capacity), and the Tower of London (i.e., planning ability) explained 27%, 39%, and 47% of the variance of system's perceived usability ratings respectively. However, the best model was the one with the number of prompts as a predictor (see Table 7 and Figure 14). The model explained 57% of the variance of the usability ratings, postulating that the individuals with ASD, who required fewer prompts for efficiently performing the VR social scenarios, perceived a higher system's usability.

**Table 7.** Linear Regression: Best Models For Predicting System's Perceived Usability.

| Predictor | F | p-value (F) | β coefficient | p-value (β) | R² |
|---|---|---|---|---|---|
| Prompts | 30.81 | < 0.001*** | -0.79 | < 0.001*** | 0.57 |
| ToL | 20.37 | < 0.001*** | 0.69 | < 0.001*** | 0.47 |
| DS Forward | 14.98 | < 0.001*** | 0.67 | < 0.001*** | 0.39 |
| Task Completion | 8.64 | 0.01** | 0.52 | 0.01** | 0.27 |

DS = Digit Span; ToL = Tower of London; *p <.05, ** p <.01, ***p <.001.

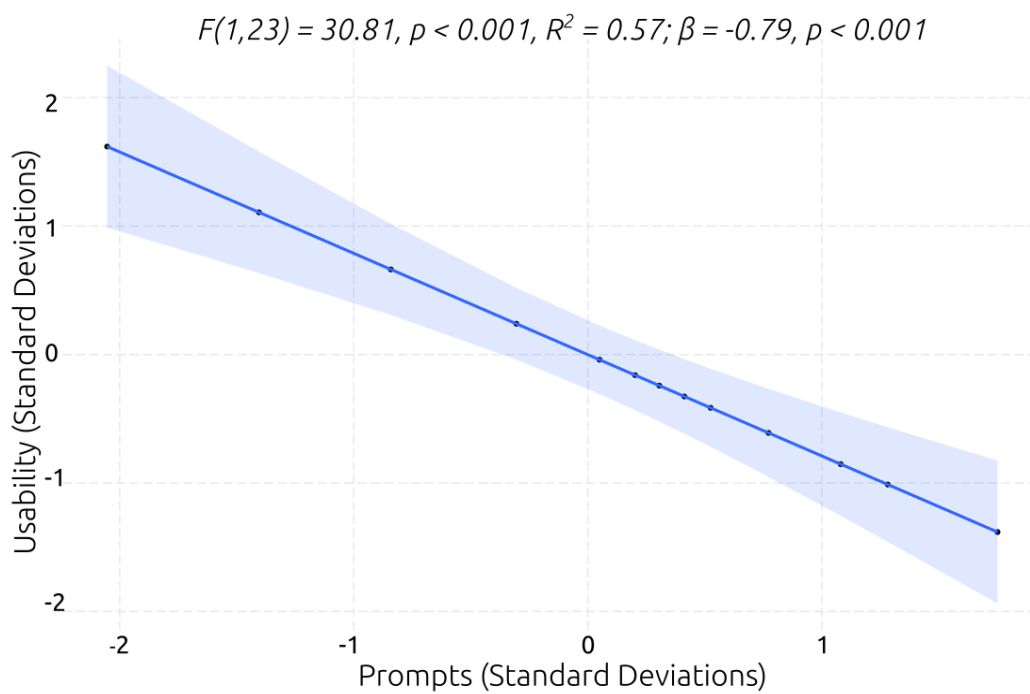

$F(1,23) = 30.81, p < 0.001, R^2 = 0.57; \beta = -0.79, p < 0.001$

**Figure 14.** Best Linear Regression Model For Predicting System's Perceived Usability.

## 4. Discussion

The present study aimed firstly to assess the usability, user experience, and acceptability of an immersive VR social skills training software (i.e., VRESS) in adults with ASD. Results showed that in terms of the system's ratings, the VRESS software exhibited a relatively high performance with positive evaluations, as average scores were close to the high edge of the possible scores of questionnaires. Secondly, the examination of the associations between mental state/emotion recognition, EF, functionality level of individuals with ASD, performance in VR social scenarios, and the self-reported ratings revealed several statistically significant relations. Furthermore, the regression models' (single predictor) analyses revealed significant predictors of several aspects. The performance in VR social scenarios (i.e., the number of prompts required to perform efficiently the social tasks) was the best significant predictor of ASD functionality level, as well as the ratings of the VR system's perceived usability, and VR social skills' training acceptability. Inhibition speed (i.e., the response time on the Stroop task) was also a significant predictor of ASD functionality level. Working memory (i.e., performance on the Digit Span Forward task) was the second-best predictor of ASD functionality level and a significant predictor of the VR system's perceived usability. Finally, the planning ability (i.e., performance on the Tower of London test) was the second-best predictor of the VR system's perceived usability and the best predictor of performance in VR social scenarios. Overall, the results of this study offered interesting insights into the utility and feasibility of VR social skills training in ASD, the possible implication of EF in social skills, and the importance of social skills in ASD severity.

### 4.1. VR Training of Social Skills in Adults with ASD

Based on the authors of the SUTAQ, UXQ, and SUS recommendations for interpreting their scores for technological interventions' acceptability [103], quality of the user experience facilitated by the software [90], [104], and system's usability [92], the VRESS showed very high acceptability, user experience, and usability, as rated by the participants with ASD. High acceptability suggests that this software [103], which also facilitates remote intervention and training of social skills, will probably be preferred by adults with ASD. Likewise, the very high usability indicates that the VR software requires a small amount of effort from the user/trainee [92]. In VRESS, the user had only to speak to the 3D characters by using the microphone of the headset and navigating in the virtual environment by pressing a single large button on the controller (either left or right). Hence, from the trainees' part, only a single button was required to be used, while the rest were controlled and operated by the researcher (see subsection 2.1. for details). Finally, the high user experience postulates that the VR software offered a highly pleasant and immersive experience to the trainee [90], [104]. Given that providing a therapeutic process, which is perceived as pleasant and positive by the patients, enhances the engagement and commitment to therapy, as well as the effect size of the therapeutic outcomes [105], the high user experience of VRESS suggests that it may achieve comparable positive outcomes.

Furthermore, given that there is a scarcity of robust evidence supporting the feasibility and acceptability of implementing immersive VR interventions in populations with ASD [59], [60], [62], the results of this study provide substantial evidence that the implementation of immersive VR social skills training in ASD is feasible and acceptable by adults with ASD. However, it should be noted that the VRESS was developed in line with the guidelines for developing VR software for psychological sciences [71], which also lead to VR software which meets the criteria of AACN and NAN [72]. For this reason, beyond the high ratings in terms of acceptability, user experience, and usability, the participants experienced minimal to absent symptoms of cybersickness, which indicates that VRESS is a VR software that meets the health and safety criteria. Finally, since the VRESS was designed specifically for individuals with ASD, the observed high ratings of acceptability, user experience, and usability, highlight the necessity of developing VR software which considers the highly prevalent cognitive and behavioural symptoms in ASD. However, a downside was that the usability and acceptability of VRESS were significantly predicted by the performance in social scenarios. This finding indicates that the negative feeling that was experienced when the participants performed poorly, influenced them to rate VRESS with lower scores, while the positive feeling of accomplishment influenced more positive scores. Both error correction and errorless learning have been seen as effective in ASD [106], however, the results of this study suggest that an errorless approach in VR social skills training may result in even higher acceptability and perceived usability. Thus, instead of receiving prompts from the operator/researcher of VRESS, which may be perceived as external corrections, the VR system may provide in-game guidance to promote an errorless completion of social tasks, while making the trainees feel that they completed them without external assistance (e.g., with the help of the researcher). Thus, an errorless approach should be preferred in a future iteration of VRESS.

### 4.2. Demographics' Role in Cognition

#### 4.2.1. Executive Functions

Results showed that verbal working memory was correlated with the participants' education. The relationship between digit span scores and education is not surprising, considering that the majority of academic tasks involve reading and lectures, which rely heavily on verbal working memory. Working memory plays an important role in educational attainment as it is consistently found to predict academic success [107], [108]. Involved in the maintenance and processing of information [109], working memory significantly associates with broad reading, comprehension and mathematical abilities [110]–[112]. In terms of inhibition, Stroop response time was shown to significantly associate with gaming experience ratings and usability. Findings of faster inhibition relating to higher perceived usability scores could suggest that the ability to suppress automatic responses/ignore distractions faster allowed participants to better use and interact with the software. The significant association between gaming experience and inhibition response time is in line with previous evidence showing that video gamers generally demonstrate faster reaction times and fewer errors relative to non-gamers [113]. Action video gamers were also found to have faster visual and auditory information processing; thus they presented faster response times than non-gamers [114]. Indeed, practising tasks which rely on inhibition and working memory -such as videogames- may lead to improved performance on similar tasks [115].

4.2.2. Mental State/Emotion Recognition

With regards to the mental state/emotion recognition ability, it was not found significantly related to performance in VR social scenarios but associated only with computing and gaming experience variables. Previous evidence suggests that individuals with ASD present difficulties in recognition of mental states/emotions (e.g., [116]–[118]), but there are limited and mixed findings regarding its association with social competence (e.g., [26]). Generally, as already discussed in the Introduction, socio-cognitive abilities (such as the recognition of mental states/emotions) do not present consistent associations with social skills in ASD. Our results showed that in adults with ASD, plausibly, it is other cognitive functions (such as EF) that are more strongly implicated in the expression of social skills. Considering though that there could be the case that no single cognitive construct can explain the whole variance of social difficulties in ASD, further research is needed to shed more light on this association. Future studies may also take into consideration other emotional and relational factors that could potentially contribute to social skills. For example, individuals with ASD may have difficulty regulating their emotions (emotional regulation) or sharing others' feelings (e.g., empathy) which can make it challenging for them to respond appropriately to social cues and situations. Accordingly, low self-esteem, negative interpersonal relationships or even low social motivation may also play a role in shaping social skills of individuals with or without ASD. Finally, the correlations between mental state/emotion recognition, gaming experience, and computing experience reveal that individuals who had more experience with video games were more able to recognize mental states/emotions in the RTMIE test. Due to their interactive nature, modern video games offer realistic cinematics and compelling avatars with complex facial expressions which may enhance gamers' ability to attribute and recognize emotions and mental states in real life contexts.

*4.2. Executive Functions and Social Skills*

Gollwitzer's *implementation intention* pertains to the formulation of an effective plan of action, which incorporates the associations between a cue with the intended action (e.g., if I encounter the X then I will do the Y) [119]. Correspondingly, the planning ability is an executive function which requires thinking about the future and respectively organising and prioritising future actions for achieving the desired goal(s) [15], [120]. In everyday life, planning defines when and where an action will take place, and updates/prioritises the plan of action based on the acquired information (e.g., I received a notification for my overdue subscription to the gym, so, I need to renew it this evening) [120]. As a result, an impaired planning ability is highly prevalent in clinical populations with reduced everyday functionality [121], [122], as well as in ASD [123], [124]. In this study, planning ability was measured by the Tower of London test, which requires individuals to generate an explicit plan of action, including all the necessary steps, towards achieving their goal [81]. Planning ability was found to be the best predictor of performance in VR social scenarios performance (i.e., the number of prompts). Comparable to everyday life, the VR social scenarios required participants to plan/implement strategies of how to move their bodies, modulate the tone of their voices, express their thoughts and perspectives, and decide to which person and how they should interact with them for achieving the respective social goals (e.g., choosing a film and buying tickets for it). Participants with ASD who presented lower planning abilities experienced more difficulties in performing the required tasks in these social scenarios. On the other hand, participants with ASD, who had better planning ability, required fewer prompts to perform the social tasks in VRESS suggesting that their planning ability has assisted them with performing social interactions without requiring support. These results and interpretation are in line with the findings of studies in children with fetal alcohol spectrum disorders [121] and 22q11 deletion syndrome [122], where planning ability was a significant predictor of social skills. Note that, comparable to individuals with ASD [123], [124], individuals with these syndromes frequently have impaired social skills and planning abilities [121], [122]. Taken together with the results of this and previous studies, planning skills are likely to facilitate social interactions as individuals need to plan and monitor their own and others' actions to adjust their responses and behaviour. Successful social interactions thus require not only the manipulation of one's and others' perspectives or the processing of social cues (i.e., working memory) but may also need planning abilities to select behavioural decisions and strategies. It should be noted at this point that social strategies may involve conscious planning as discussed above, but of course social behaviours may also manifest unconsciously (particularly in everyday life) as they are often based on previous interactions or emotional experiences.

In line with a review of studies on working memory impairments in ASD, where lower scores in verbal working memory were associated with greater problems in adaptive behaviour [125], in this study, verbal working memory was correlated with the performance in VR social scenarios (i.e., the number of prompts). Performance in situations such as the social interactions presented in VRESS scenarios places high demands on processing which in turn demands increased controlled attentional processing by the executive system of working memory. Participants with ASD, who had higher digit span scores, required fewer prompts to perform the social tasks in VRESS, suggesting that working memory may facilitate social interactions without individuals needing support and/or reminders. Cognitive structures such as the recognition and understanding of others' thoughts, beliefs and mental/emotional states during social interactions require a heavier load of working memory [126] as individuals have to actively maintain and manipulate personal perspectives and new, complex information of external social cues. Accordingly, social interactions could be considered as a dual task (i.e., based on one having to balance personal perspectives with those of the people interacting with) and for that reason require working memory mechanisms [127]. Taking all these together, it is likely that participants with ASD who have lower working memory abilities required more prompts to complete the social scenarios because effective social cognition and social interaction are not possible unless one can maintain and process perspectives, social cues, and communicative strategies effectively. Nevertheless, working memory ability was not a significant predictor of performance in VR social scenarios, suggesting that its implication in social skills may be secondary and/or moderating. Indeed, this interpretation of our findings agrees with the findings of a recent study, where a moderating role of working memory, between verbal ability and social skills, was observed during early schooling years where the acquisition of social skills is crucial [128].

### 4.3. Predictors of Functionality Level in ASD

Our results indicated that the ASD functionality level was related to and predicted by inhibition and verbal working memory supporting previous evidence that has pinpointed a link between EF and the later severity features/symptoms in ASD [129]. Generally, impaired EF have been proposed to underlie core severity symptoms of the spectrum [125], [130]. In line with this evidence, our results suggest that executive functions are central to ASD and highlight their importance as a crucial domain for support and training/intervention. It should be noted at this point though that less attention has been generally given to the examination of potential cognitive factors which may be crucial for the implementation of timely and effective interventions in ASD. Future longitudinal studies can further clarify whether executive functions have prognostic significance in adults with ASD.

Most importantly though, the performance in VR social scenarios (i.e., number of prompts) was found to be the best significant predictor of ASD functionality level. Impaired social and communication skills are core features of ASD, which is common across the spectrum regardless of the functionality level [131]–[134]. Although some of the best predictors of ASD severity/functionality in childhood are language level [135] or IQ [136], the severity of social and communication skills have been found to associate with [131], [134], or differ across    [132], [133], the diverse functionality levels within ASD diagnosis. Observing the performance in VR social scenarios as a significant predictor of ASD functionality level is thus aligned with the findings of the aforementioned studies. However, it should be noted that the results of this study indicated that social skills were not just a significant predictor, but the best predictor of functionality level in ASD. Given that the participants of this study were either diagnosed with functionality levels 1 and 2 (i.e., high and moderate functioning respectively) based on DSM-5 [1], this outcome postulates that social skills may potentially serve as a central indicating factor of functionality in high and moderate functioning adults with ASD. Notably, the social scenarios of VRESS benefit from enhanced ecological validity, which allows the depiction of everyday functionality [52], [72]. Thus, this outcome may be also attributed to the enhanced ecological validity of VRESS social scenarios, which encompass the complexity and the demands of social contexts and situations in daily life.

### 4.4. Limitations and Future Studies

The findings of the present study should be interpreted considering its limitations. The present sample of adults with ASD may not represent the more general population of the spectrum. Participants' average age was approximately 30 years, being mostly representative of early adulthood (i.e., 20-39 years old). Future studies are thus required to establish whether these results can be replicated across younger children, adolescents, and/or older adults. Furthermore, the current study was not a randomised controlled trial (RCT) study to effectively examine the efficiency of VRESS in improving the social skills of individuals with ASD. Future studies should hence consider conducting an RCT experimental protocol, also incorporating a control group, to scrutinize the efficiency of the VR interventions in enhancing the social skills of adults with ASD. Finally, the VRESS did not offer an errorless learning approach, which our results showed may be beneficial for adults with ASD. Future iterations of VRESS should facilitate an errorless learning approach for improving its efficacy. Finally, future VR studies are needed to identify more potential prognostic markers of cognitive and social functioning in ASD.

### 5. Conclusions

The VRESS appears an appropriate VR social skills training system, which facilitates a high acceptability, usability, and user experience in individuals with ASD, without inducing adverse symptoms. These positive outcomes pertaining to VRESS also support the effectiveness and feasibility of implementing VR social skills training in individuals with ASD. Furthermore, executive functions were found to be implicated in the social skills of adults with ASD. Finally, social skills were seen as the best indicator of the severity/functionality level of adults with ASD.

**Author Contributions:** Conceptualization, K.P., V.M., C.S., and A.P methodology, P.K., E.K., C.S., and A.P; software, P.K. and C.S.; validation, P.K., E.K., P.R., V.M., K.P., C.S., and A.P; formal analysis, P.K. and P.R.; investigation, P.K. and E.K.; resources, K.P., C.S., V.M., and A.P; data curation, P.K., E.K., and P.R.; writing—original draft preparation, P.K. and E.K.; writing—review and editing, P.K., E.K., P.R., V.M., K.P., C.S., and A.P; visualization, P.K. and P.R.; supervision, P.R., K.P., and A.P; project administration, P.K. and C.S.; funding acquisition, C.S.. All authors have read and agreed to the published version of the manuscript.

**Funding:** The VRESS project is co-financed by the European Regional Development Fund of the European Union and Greek national funds through the Operational Program Competitiveness, Entrepreneurship, and Innovation, under the call RESEARCH – CREATE – INNOVATE (project code: T1EDK-01248).

**Institutional Review Board Statement:** The study was conducted in accordance with the Declaration of Helsinki and approved by the Ethics Committee of Eginition Hospital (117/16.03.2020).

**Informed Consent Statement:** Informed consent was obtained from all subjects involved in the study.

**Data Availability Statement:** The data presented in this study are available on request from the corresponding author. The data are not publicly available due to ethical approval requirements.

**Acknowledgements:** We would like to thank the participants for their attendance and commitment to this research project. Also, we deeply thank Omega Technology for developing VRESS and allowing us to use it in this study.

**Conflicts of Interest:** The authors declare no conflict of interest.